\documentclass[%
 reprint,
 amsmath,amssymb,
 aps,
pra,
]{revtex4-2}

\usepackage{xcolor}
\usepackage{graphicx}
\usepackage{dcolumn}
\usepackage{bm}
\usepackage{gensymb}

\begin{document}


\title{\Large Spin-Flip Configuration Interaction for Strong Static Correlation in Quantum Electrodynamics}

\author{Braden M. Weight}
\email{braden.m.weight@lanl.gov}
\affiliation{Theoretical Division, Center for Integrated Nanotechnologies, Los Alamos National Laboratory, Los Alamos, NM, 87545, U.S.A.}

\author{Zheng Pei}
\affiliation{\small Department of Chemistry, Brandeis University, Waltham, MA 02453, U.S.A. }



\author{Sergei Tretiak}
\email{serg@lanl.gov}
\affiliation{Theoretical Division, Center for Integrated Nanotechnologies, Los Alamos National Laboratory, Los Alamos, NM, 87545, U.S.A.}

\date{\today}

\maketitle

In computational chemistry of molecular materials, strong static correlation effects appear when electronic states, often involving the ground state, become quasi-degenerate, as occurs, for example, in bond-breaking processes. Such situations present significant challenges for accurate theoretical treatment. In these regimes, many-body methods involving a single-determinant description, such as Hartree-Fock theory and its time-dependent extension, fail to reproduce the correct topology of the ground and excited state potential energy surfaces (\textit{e.g.}, near conical intersections). When strongly correlated electronic systems are further strongly coupled to a quantized radiation field within the framework of non-relativistic cavity quantum electrodynamics, an additional photonic degree of freedom introduces both new complexity and new opportunities to control. Excited cavity photons can modify bond-breaking processes and enable tunability of geometrical and spin-phase transitions, for instance, in organometallic complexes. To overcome this bottleneck, in this work, we extend the well-studied spin-flip configuration interaction singles (SF-CIS) approach to explicitly include quantized cavity photons leading to QED-SF-CIS method. We derive the spin-flip Hamiltonian and find that the double excitation subspace of the system (single with respect to electronic excitation) must be included in the configurations to properly describe singlet electronic states interacting with cavity photons. We then illustrate, through representative molecular examples, how cavity coupling can provide additional tunability in bond-breaking processes. We finally generalize this approach to include higher numbers of photonic excitations, which are required in the strong coupling regime.

\section{Introduction}
Treatment of strongly correlated systems has long been at the forefront of molecular and solid state science. While such systems give rise to rich and novel physical phenomena, paradigmatic models, $e.g.$, Hubbard model and similar, are notoriously difficult to solve. Even for a moderate number of degrees of freedom, exact solutions quickly become intractable. For realistic systems containing 100s or 1000s of electrons, no attempt can be made to solve them exactly. This situation is further complicated when strongly interacting matter is coupled to a quantized electromagnetic field under strong light–matter coupling conditions, forming hybrid light–matter states known as polaritons.  The inclusion of photonic degrees of freedom also requires strict quantum mechanical treatment and could yield increased (or decreased) electron-electron correlation mediated by the photonic degree of freedom.\cite{mandal_theoretical_2023,weight_theory_2023,ruggenthaler_understanding_2023,weight_ab_2025-1}

Electronic correlation in molecules is commonly classified into two well-accepted types, often referred to as static and dynamic correlations. In the case of static correlation (the main focus of this work), the correlation stems from nearly degenerate electronic configurations. Common examples for this type of correlation are bond dissociation models, \textit{e.g.}, H$_2$, LiF, F$_2$, Cr$_2$, etc., rotations, \textit{e.g.}, ethylene dihedral twisting, and bond formation in chemical reactions, \textit{e.g.}, Diels-Alder reactions. There are many tools available for simulating strongly correlated materials. The most obvious solution is to use multi-reference methods, such as MC-SCF (\textit{e.g.}, CASSCF) and its perturbative extensions (\textit{e.g.}, CASPT2). However, even for nominally single-reference methods, there are strategies to treat strong static correlation. One such approach is the so-called spin-flip method,\cite{krylov_spin-flip_2001,shao_spinflip_2003,casanova_spin-flip_2020} where the reference is calculated at an adjacent multiplicity (\textit{e.g.}, triplet) and spin-flipping excitations are performed to obtain a different multiplicity (\textit{e.g.}, singlet). This method has proven to be powerful in overcoming static correlation that is mediated by nearly degenerate electronic configurations of the same multiplicity because the ground and excited states are treated on the same footing (\textit{i.e.}, as 
excited states of a different reference). Therefore, the topology of the ground and excited states is immediately corrected.  This feature is especially important in nonadiabatic dynamics, where conical intersections of S$_0$/S$_1$ often pose unique challenges for simulation of no-radiative relaxation due to the incorrect topology near these conical intersections.

\section{Methods}
\subsection{Spin-Flip Configuration Interaction Singles (SF-CIS)}

In the realm of single reference methods for calculating electronically excited states, the CIS approach is an approximation of the full time-dependent Hartree-Fock (TDHF) equations in that the de-excitation matrix, often denoted as $\boldsymbol{B}$, is neglected. The TDHF equations may relate to the first order random phase approximation (RPA). For time-dependent density functional theory (TDDFT)-based response methods, the corresponding nomenclature is the Tamm-Dancoff Approximation (TDA), which serves as the TDDFT analogue of CIS. In either case, it is generally well-accepted that the CIS for TDHF (or TDA for TDDFT) is often preferable for a variety of practical reasons, such as reduced numerical cost, the alleviation of the triplet instability problem and real eigenvalues via the use of a Hermitian eigenvalue problem.\cite{casida_progress_2012,xu_combined_2013}

The full CIS Hamiltonian can be written in spin-space matrix blocks as,
\begin{align}\label{EQ:H_FULL_CIS}
    &\hat{H}_\mathrm{CIS}=\begin{bmatrix}
        0&0&0&0&0\\
        0&{\bf A}_{\alpha\alpha,\alpha\alpha}&{\bf A}_{\alpha\alpha,\beta\beta}&0&0\\
        0&{\bf A}_{\beta\beta,\alpha\alpha}&{\bf A}_{\alpha\alpha,\alpha\alpha}&0&0\\
        0&0&0&{\bf A}_{\alpha\beta,\alpha\beta}&0\\
        0&0&0&0&{\bf A}_{\beta\alpha,\beta\alpha}
    \end{bmatrix}
\end{align}
where the reference energy has already been subtracted from the diagonal elements as $\hat{H}_\mathrm{CIS} - E_\mathrm{HF} \rightarrow \hat{H}_\mathrm{CIS}$. From left to right (or top to bottom), the basis configurations (in spin blocks) of the matrix are written as $|\mathrm{HF}\rangle$, $|\Phi^\alpha_\alpha\rangle$, $|\Phi^\beta_\beta\rangle$, $|\Phi^\alpha_\beta\rangle$, $|\Phi^\beta_\alpha\rangle$, where $|\Phi_{i_\alpha}^{a_\beta}\rangle \equiv \hat{c}^\dagger_{a_\beta}\hat{c}_{i_\alpha}|\mathrm{HF}\rangle$ is a Slater determinant constructed by a single substitution on the HF reference determinant of occupied orbital $i_\alpha$ with spin $\alpha$ by a virtual orbital $a$ of spin $\beta$, denoted by the fermionic creation ($\hat{c}^\dagger_{a_\beta})$ and annihilation ($\hat{c}_{i_\alpha}$) operators.

The subsequent matrix elements (in chemist's notation) of each block are written as,
\begin{align}
    &A_{i_\alpha a_\alpha,j_\alpha b_\alpha} = (\epsilon_{a_\alpha} - \epsilon_{i_\alpha})\delta_{i_\alpha j_\alpha }\delta_{a_\alpha b_\alpha }\nonumber\\ 
    &~~~~~~~~~~~~~~~~+ (i_\alpha a_\alpha |j_\alpha b_\alpha ) - (i_\alpha j_\alpha |a_\alpha b_\alpha )
\end{align}
\begin{align}
    &A_{i_\beta a_\beta,j_\beta b_\beta} =(\epsilon_{a_\beta} - \epsilon_{i_\beta})\delta_{i_\beta j_\beta}\delta_{a_\beta b_\beta }\nonumber\\ 
    &~~~~~~~~~~~~~~~~+ (i_\beta a_\beta |j_\beta b_\beta ) - (i_\beta j_\beta |a_\beta b_\beta )
\end{align}
\begin{align}
    A_{i_\alpha a_\alpha,j_\beta b_\beta} = (i_\alpha a_\alpha |j_\beta b_\beta )
\end{align}
\begin{align}
    A_{i_\beta a_\beta,j_\alpha b_\alpha} = (i_\beta a_\beta |j_\alpha b_\alpha )
\end{align}
\begin{align}
    &A_{i_\alpha a_\beta,j_\alpha b_\beta} =(\epsilon_{a_\beta} - \epsilon_{i_\alpha})\delta_{i_\alpha j_\alpha}\delta_{a_\beta b_\beta }\nonumber\\
    &~~~~~~~~~~~~~~~~~~~~~~- (i_\alpha j_\alpha |a_\beta b_\beta )
\end{align}
\begin{align}
    &A_{i_\beta a_\alpha,j_\beta b_\alpha} =(\epsilon_{a_\alpha} - \epsilon_{i_\beta})\delta_{i_\beta j_\beta}\delta_{a_\alpha b_\alpha }\nonumber\\
    &~~~~~~~~~~~~~~~~~~~~~~- (i_\beta j_\beta |a_\alpha b_\alpha )
\end{align}
Note that in the CIS approximation of TDHF, the spin-flipping blocks ${\bf A}_{\alpha\beta,\alpha\beta}$ and ${\bf A}_{\beta\alpha,\beta\alpha}$ are decoupled.\cite{hait_beyond_2019} Thus, this matrix block can be constructed and solved independently of all other blocks. Thus, the SF-CIS Hamiltonian is simply,
\begin{equation}\label{EQ:H_ab_SF_BLOCK}
    \hat{H}^\mathrm{SF-CIS}_{\alpha\rightarrow \beta} = {\bf A}_{\alpha\beta,\alpha\beta}
\end{equation}
or
\begin{equation}\label{EQ:H_ba_SF_BLOCK}
    \hat{H}^\mathrm{SF-CIS}_{\beta\rightarrow \alpha} = {\bf A}_{\beta\alpha,\beta\alpha}.
\end{equation}

\subsection{ Quantum Electrodynamics Hartree-Fock (QED-HF) }

The non-relativistic Pauli-Fierz Hamiltonian can be written as,
\begin{equation}\label{EQ:H_PF}
    \hat{H}_\mathrm{PF} = \hat{H}_\mathrm{el} + \omega_\mathrm{c} \hat{a}^\dagger\hat{a} + \sqrt{\frac{\omega_\mathrm{c}}{2}}\vec{\lambda}\cdot\vec{\hat{d}}(\hat{a}^\dagger + \hat{a}) + \frac{1}{2} (\vec{\lambda} \cdot \vec{\hat{d}})^2,
\end{equation}
where $\hat{H}_\mathrm{el}$ is the standard electronic Hamiltonian, composed of the one- and two-body terms, in the Born-Oppenheimer approximation. The light-matter coupling strength is defined as
\begin{equation}
    \vec{\lambda} = \sqrt{\frac{1}{\epsilon\mathcal{V}}}~\vec{e},
\end{equation}
where $\epsilon$ is the dielectric constant of the media, $\mathcal{V}$ is the cavity mode volume, and $\vec{e}$ is the electric polarization unit vector of the cavity mode.

Often, one chooses the the coherent state basis for the cavity photons\cite{haugland_coupled_2020}, which shifts the molecular dipole operator by a constant via unitary transformation of the Hamiltonian with $\hat{U}^\mathrm{CS} = e^{-\frac{\vec{\lambda}x}{\sqrt{2\omega_\mathrm{c}}}\cdot\langle \vec{\hat{d}} \rangle (\hat{a}^\dagger + \hat{a})}$, taken to be the expectation value of the molecular dipole in the reference state (\textit{i.e.}, often HF or DFT ground state), resulting in
\begin{equation}\label{EQ:H_PF_CS}
    \hat{H}_\mathrm{PF}^\mathrm{CS} = \hat{H}_\mathrm{el} + \omega_\mathrm{c} \hat{a}^\dagger\hat{a} + \sqrt{\frac{\omega_\mathrm{c}}{2}}\vec{\lambda}_\mathrm{c}\cdot\vec \Delta{\hat{d}}(\hat{a}^\dagger + \hat{a}) + \frac{1}{2}\vec{\lambda}_\mathrm{c}^2 \cdot (\Delta \vec{\hat{d}})^2,
\end{equation}
where $\Delta \vec{\hat{d}} = \vec{\hat{d}} - \langle \vec{\hat{d}} \rangle_\mathrm{HF}$ with $\langle \vec{\hat{d}} \rangle_\mathrm{HF} \equiv \langle \mathrm{HF} | \vec{\hat{d}} | \mathrm{HF} \rangle$. It should be noted that more sophisticated representations for the photonic degrees of freedom have been developed.\cite{li_first-principles_2023,mazin_light-matter_2024,mandal_polarized_2020}

In the case of HF theory, the energy is then calculated by expectation value of a single determinant composed of the tensor product of the Slater determinant and the cavity vacuum state as,
\begin{align}
    E_\mathrm{QED-HF}^\mathrm{CS} &= \langle \mathrm{HF}, 0 |  \hat{H}_\mathrm{PF}^\mathrm{CS} | \mathrm{HF}, 0 \rangle\nonumber\\
    &= \langle \mathrm{HF}|  \hat{H}_\mathrm{el} | \mathrm{HF}\rangle \langle 0 | 0\rangle + \langle 0| \omega_\mathrm{c}\hat{a}^\dagger\hat{a}|0 \rangle \nonumber\\
    &+ \sqrt{\frac{\omega_\mathrm{c}}{2}} \vec{\lambda}\cdot\langle \mathrm{HF} | \Delta \vec{\hat{d}} | \mathrm{HF} \rangle \langle 0 | \hat{a}^\dagger + \hat{a} | 0 \rangle \nonumber\\ 
    &+\frac{1}{2}\vec{\lambda}\cdot\langle \mathrm{HF} | (\Delta\vec{\hat{d}})^2 | \mathrm{HF} \rangle \cdot \vec{\lambda} ~\langle 0 | 0 \rangle \nonumber\\
    &= E_\mathrm{HF} + E^\mathrm{QED}_\mathrm{RHF}.
\end{align}
with $E^\mathrm{QED}_\mathrm{RHF} = \frac{1}{2}\vec{\lambda}\cdot\langle \mathrm{HF} | (\Delta\vec{\hat{d}})^2 | \mathrm{HF} \rangle \cdot \vec{\lambda}$ is the residual dipole self-energy (DSE) term after the coherent state transformation. Note that the bilinear term does not explicitly appear in the coherent state transformation within HF since $\langle \mathrm{HF} |  \vec{\hat{d}} - \langle\vec{\hat{d}}\rangle_\mathrm{HF} | \mathrm{HF} \rangle = 0$. To achieve this result, the orthogonality of the cavity number states, $\langle n | m \rangle = \delta_{nm}$ as well as the orthonormality of the Slater determinant, $\langle \mathrm{HF} | \mathrm{HF} \rangle = 1$, was used.

For the case of spin-restricted HF (RHF), the bare RHF energy is decomposed in the usual way as,
\begin{equation}\label{EQ:E_RHF}
    E_\mathrm{RHF} = \sum_{\mu \nu} (2h_{\mu\nu} + 2J_{\mu\nu} - K_{\mu\nu}) \rho_{\mu\nu}
\end{equation}
where $\rho_{\mu\nu} = \sum_i C_{\mu i}^* C_{\nu i}$ is the one-particle density matrix in the atomic orbital representation $\{\mu,\nu\}$ and $C_{\mu i}$ is the $\mu_\mathrm{th}$ atomic orbital coefficient of the $i_\mathrm{th}$ molecular orbital. $h_{\mu\nu}$ is a single-election part of the Fock matrix, whereas the Coulomb ($J$) and exchange ($K$) integrals are defined as
\begin{equation}\label{EQ:J}
    J_\mathrm{\mu\nu} = \sum_{\xi \delta} (\mu\nu|\xi\delta)\rho_{\xi\delta}
\end{equation}
and
\begin{equation}\label{EQ:K}
    K_\mathrm{\mu\nu} = \sum_{\xi \delta} (\mu\delta|\xi\nu)\rho_{\xi\delta}.
\end{equation}

The DSE term is similarly expanded in the one- and two-particle contributions as,
\begin{equation}\label{EQ:E_QED_RHF}
    E^\mathrm{QED}_\mathrm{RHF} = \sum_{\mu \nu} (2h^\mathrm{QED}_{\mu\nu} + 2J^\mathrm{QED}_{\mu\nu} - K^\mathrm{QED}_{\mu\nu}) \rho_{\mu\nu},
\end{equation}
with
\begin{align}
    h^\mathrm{QED}_{\mu\nu} &= -\vec{\lambda}\cdot\langle \vec{\hat{d}} \rangle_\mathrm{HF} ~(\vec{\lambda} \cdot \vec{\hat{d}}_{\mu\nu})\nonumber\\ 
    &~~~+ \frac{1}{2} \sum_{x_i x_j}^3 \lambda_{x_i} \cdot (Q_{x_i x_j})_{\mu\nu} \cdot \lambda_{x_j},
\end{align}
\begin{equation}\label{EQ:J_QED}
    J^\mathrm{QED}_{\mu\nu} = \sum_{\sigma \rho} \rho_{\sigma \rho} (\vec{\lambda} \cdot \vec{d}_{\mu\nu}) (\vec{\lambda} \cdot \vec{d}_{\sigma\rho}),
\end{equation}
and
\begin{equation}\label{EQ:K_QED}
    K^\mathrm{QED}_{\mu\nu} = \sum_{\sigma \rho} \rho_{\sigma \rho} (\vec{\lambda} \cdot \vec{d}_{\mu\rho}) (\vec{\lambda} \cdot \vec{d}_{\sigma\nu}).
\end{equation}
Note that $(\vec{\lambda}\cdot\Delta \vec{\hat{d}})^2 = [\vec{\lambda}\cdot(\vec{\hat{d}} - \langle \vec{\hat{d}}\rangle_\mathrm{HF})]^2 = (\vec{\lambda}\cdot\vec{\hat{d}})^2 + 2(\vec{\lambda}\cdot\langle \vec{\hat{d}}\rangle_\mathrm{HF})(\vec{\lambda}\cdot\vec{\hat{d}}) + (\vec{\lambda}\cdot\langle \vec{\hat{d}}\rangle_\mathrm{HF})^2$, where the first term is a two-electron operator appearing in $J^\mathrm{QED}_{\mu\nu}$ and $K^\mathrm{QED}_{\mu\nu}$, the second term is a one-electron operator appearing in $h^\mathrm{QED}_{\mu\nu}$, and the third term is a constant.

For the unrestricted case, the energy functional is instead
\begin{equation}\label{EQ:E_UHF}
    E_\mathrm{UHF} = \sum_{\sigma} \sum_{\mu \nu} \bigg(h_{\mu\nu} + \frac{1}{2}\bigg[J_{\mu\nu,\alpha} + J_{\mu\nu,\beta} - K_{\mu\nu,\sigma}\bigg]\bigg) \rho_{\mu\nu,\sigma}
\end{equation}
and
\begin{equation}\label{EQ:E_QED_UHF}
    E^\mathrm{QED}_\mathrm{UHF} = \sum_{\sigma} \sum_{\mu \nu} \bigg(h^\mathrm{QED}_{\mu\nu} + \frac{1}{2}\bigg[J^\mathrm{QED}_{\mu\nu,\alpha} + J^\mathrm{QED}_{\mu\nu,\beta} - K^\mathrm{QED}_{\mu\nu,\sigma}\bigg]\bigg) \rho_{\mu\nu,\sigma},
\end{equation}
where the integrals $J$, $K$ in Eqns.~\ref{EQ:J},~\ref{EQ:K},~\ref{EQ:J_QED}, and~\ref{EQ:K_QED} are contracted with the spin-specific one-particle density matrix, $\rho_{\mu\nu,\sigma} = C^*_{\mu i,\sigma} C_{\nu i,\sigma}$.

\subsection{QED-SF-CIS}

This QED-SF-CIS Hamiltonian can require an extended basis of determinants to access the cavity interactions with the singlet manifold. These quantities, from left to right in Eq.~\ref{EQ:H_QED_FULL_CIS} below, are $|\mathrm{HF}, 0\rangle$, $|\Phi^\alpha_\alpha, 0\rangle$, $|\Phi^\beta_\beta, 0\rangle$, $|\Phi^\alpha_\beta, 0\rangle$, $|\Phi^\beta_\alpha, 0\rangle$, $|\mathrm{HF}, 1\rangle$, $|\Phi^\alpha_\alpha, 1\rangle$, $|\Phi^\beta_\beta, 1\rangle$, $|\Phi^\alpha_\beta, 1\rangle$, and $|\Phi^\beta_\alpha, 1\rangle$, where we assume a tensor product of electronically excited Slater determinants and excitations of cavity photons as $|\Phi^\alpha_\alpha\rangle \otimes |1\rangle \equiv |\Phi^\alpha_\alpha, 1\rangle$.

Similarly to Eq.~\ref{EQ:H_FULL_CIS}, we write the full CIS matrix in the extended basis as described above for the PF Hamiltonian in Eq.~\ref{EQ:H_PF_CS} as,
{\tiny
\begin{widetext}
    \begin{align}\label{EQ:H_QED_FULL_CIS}
        &\hat{H}_\mathrm{QED-CIS}=\nonumber\\
        &\begin{bmatrix}
            0&0&0&0&0&{\color{red}\langle \mathrm{HF} | \hat{d} | \mathrm{HF}\rangle}&{\color{red}\langle  \mathrm{HF} |\hat{d}| \Phi^\alpha_\alpha \rangle}&{\color{red}\langle  \mathrm{HF} |\hat{d}| \Phi^\beta_\beta \rangle}&{\color{red}0}&{\color{red}0}\\
            0&\langle \Phi_\alpha^\alpha|{\bf A}'| \Phi_\alpha^\alpha \rangle&\langle \Phi_\alpha^\alpha|{\bf A}'| \Phi_\beta^\beta \rangle&0&0&{\color{red}\langle \Phi^\alpha_\alpha |\hat{d}| \mathrm{HF} \rangle}&{\color{red}\langle \Phi^\alpha_\alpha |\hat{d}| \Phi^\alpha_\alpha \rangle}&{\color{red}\langle \Phi^\alpha_\alpha |\hat{d}| \Phi^\beta_\beta \rangle}&{\color{red}0}&{\color{red}0}\\
            0&\langle \Phi_\beta^\beta|{\bf A}'| \Phi_\alpha^\alpha\rangle&\langle \Phi_\beta^\beta|{\bf A}'| \Phi_\beta^\beta \rangle&0&0&{\color{red}\langle \Phi^\beta_\beta |\hat{d}| \mathrm{HF} \rangle}&{\color{red}\langle \Phi^\beta_\beta |\hat{d}| \Phi^\alpha_\alpha \rangle}&{\color{red}\langle \Phi^\beta_\beta |\hat{d}| \Phi^\beta_\beta \rangle}&{\color{red}0}&{\color{red}0}\\
            0&0&0&\langle \Phi_\alpha^\beta|{\bf A}'| \Phi_\alpha^\beta \rangle&0&{\color{red}0}&{\color{red}0}&{\color{red}0}&{\color{red}\langle \Phi^\beta_\alpha |\hat{d}| \Phi^\beta_\alpha \rangle}&{\color{red}0}\\
            0&0&0&0&\langle \Phi_\beta^\alpha|{\bf A}'| \Phi_\beta^\beta \rangle&{\color{red}0}&{\color{red}0}&{\color{red}0}&{\color{red}0}&{\color{red}\langle \Phi^\alpha_\beta |\hat{d}| \Phi^\alpha_\beta \rangle}\\
            {\color{red}\langle \mathrm{HF} |\hat{d}| \mathrm{HF} \rangle}&{\color{red}\langle \mathrm{HF} |\hat{d}| \Phi^\alpha_\alpha \rangle}&{\color{red}\langle \mathrm{HF} |\hat{d}| \Phi^\beta_\beta \rangle}&{\color{red}0}&{\color{red}0}&{\color{blue}\omega_\mathrm{c}}&{\color{blue}0}&{\color{blue}0}&{\color{blue}0}&{\color{blue}0}\\
            {\color{red}\langle \Phi^\alpha_\alpha |\hat{d}| \mathrm{HF} \rangle}&{\color{red}\langle \Phi^\alpha_\alpha |\hat{d}| \Phi^\alpha_\alpha \rangle}&{\color{red}\langle \Phi^\alpha_\alpha |\hat{d}| \Phi^\beta_\beta \rangle}&{\color{red}0}&{\color{red}0}&{\color{blue}0}&{\color{blue}\langle \Phi_\alpha^\alpha|{\bf A}'| \Phi_\alpha^\alpha \rangle+\omega_\mathrm{c}}&{\color{blue}\langle \Phi_\alpha^\alpha|{\bf A}'| \Phi_\beta^\beta \rangle}&{\color{blue}0}&{\color{blue}0}\\
            {\color{red}\langle \Phi^\beta_\beta |\hat{d}| \mathrm{HF} \rangle}&{\color{red}\langle \Phi^\beta_\beta |\hat{d}| \Phi^\alpha_\alpha \rangle}&{\color{red}\langle \Phi^\beta_\beta |\hat{d}| \Phi^\beta_\beta \rangle}&{\color{red}0}&{\color{red}0}&{\color{blue}0}&{\color{blue}\langle \Phi_\beta^\beta|{\bf A}'| \Phi_\alpha^\alpha\rangle}&{\color{blue}\langle \Phi_\beta^\beta|{\bf A}'| \Phi_\beta^\beta \rangle+\omega_\mathrm{c}}&{\color{blue}0}&{\color{blue}0}\\
            {\color{red}0}&{\color{red}0}&{\color{red}0}&{\color{red}\langle \Phi^\beta_\alpha |\hat{d}| \Phi^\beta_\alpha \rangle}&{\color{red}0}&{\color{blue}0}&{\color{blue}0}&{\color{blue}0}&{\color{blue}\langle \Phi_\alpha^\beta|{\bf A}'| \Phi_\alpha^\beta \rangle+\omega_\mathrm{c}}&{\color{blue}0}\\
            {\color{red}0}&{\color{red}0}&{\color{red}0}&{\color{red}0}&{\color{red}\langle \Phi^\alpha_\beta |\hat{d}| \Phi^\alpha_\beta \rangle}&{\color{blue}0}&{\color{blue}0}&{\color{blue}0}&{\color{blue}0}&{\color{blue}\langle \Phi_\beta^\alpha|{\bf A}'| \Phi_\beta^\alpha \rangle+\omega_\mathrm{c}}\\
        \end{bmatrix}
    \end{align}
\end{widetext}
}
where the zero-photon block is in black, the one-photon block is in blue, and the zero-to-one photon coupling block is in red. Note that the coupling terms are written in shorthand notation where $\langle \Phi_\alpha^\beta | \vec{\boldsymbol{\hat{d}}} | \Phi_\alpha^\beta \rangle \equiv \sqrt{\frac{\omega_\mathrm{c}}{2}}\vec{\lambda}\cdot\langle \Phi_\alpha^\beta | \Delta\vec{\boldsymbol{\hat{d}}} | \Phi_\alpha^\beta \rangle$ so that the matrix fits on the page.

Further, ${\bf A}' \equiv {\bf A} + {\bf \Delta}$ are the DSE-dressed electronic terms with
\begin{align}
    \Delta_{i_\sigma a_{\sigma'}, j_\sigma b_{\sigma'}} &= (\vec{\lambda} \cdot \Delta\vec{d}_{i_\sigma a_{\sigma'}}) (\vec{\lambda} \cdot \Delta\vec{d}_{j_\sigma b_{\sigma'}})\delta_{\sigma{\sigma'}}\nonumber\\
    &- (\vec{\lambda} \cdot \Delta\vec{d}_{i_\sigma j_\sigma}) (\vec{\lambda} \cdot \Delta\vec{d}_{a_{\sigma'} b_{\sigma'}})
\end{align}
where the first term is the Coulomb and the second term is the exchange.

The SF-blocks $\{\langle \Phi_\alpha^\beta|\hat{H}|\Phi_\alpha^\beta\rangle\}$ and $\{\langle \Phi_\beta^\alpha|\hat{H}|\Phi_\beta^\alpha\rangle\}$ of Eq.~\ref{EQ:H_QED_FULL_CIS} can be solved independently from non-spin-flipping blocks (in CIS or TDHF) as well as from one another (only in CIS) and written as,
\begin{align}\label{EQ:H_QED_ab_SF_BLOCK}
    &\hat{H}^\mathrm{QED-SF-CIS}_{\alpha\rightarrow\beta} =\nonumber\\
    &\begin{bmatrix}
        \langle \Phi_\alpha^\beta | {\bf A}' | \Phi_\alpha^\beta \rangle& \sqrt{\frac{\omega_\mathrm{c}}{2}}\vec{\lambda}\cdot\langle \Phi_\alpha^\beta | \Delta\vec{\boldsymbol{\hat{d}}} | \Phi_\alpha^\beta \rangle\\
        \sqrt{\frac{\omega_\mathrm{c}}{2}}\vec{\lambda}\cdot\langle \Phi_\alpha^\beta | \Delta\vec{\boldsymbol{\hat{d}}} | \Phi_\alpha^\beta \rangle&\langle \Phi_\alpha^\beta | {\bf A}' | \Phi_\alpha^\beta \rangle + \omega_\mathrm{c}
    \end{bmatrix}
\end{align}
or
\begin{align}\label{EQ:H_QED_ba_SF_BLOCK}
    &\hat{H}^\mathrm{QED-SF-CIS}_{\beta\rightarrow\alpha} =\nonumber\\
    &\begin{bmatrix}
        \langle \Phi_\beta^\alpha | {\bf A}' | \Phi_\beta^\alpha \rangle& \sqrt{\frac{\omega_\mathrm{c}}{2}}\vec{\lambda}\cdot\langle \Phi_\beta^\alpha | \Delta\vec{\boldsymbol{\hat{d}}} | \Phi_\beta^\alpha \rangle\\
        \sqrt{\frac{\omega_\mathrm{c}}{2}}\vec{\lambda}\cdot\langle \Phi_\beta^\alpha | \Delta\vec{\boldsymbol{\hat{d}}} | \Phi_\beta^\alpha\rangle&\langle \Phi_\beta^\alpha | {\bf A}' | \Phi_\beta^\alpha \rangle + \omega_\mathrm{c}
    \end{bmatrix}.
\end{align}

Starting from a $\alpha\alpha$-triplet reference (M$_\mathrm{s}$=+1), then solving the $\hat{H}^\mathrm{QED-SF-CIS}_{\alpha\rightarrow\beta}$ Hamiltonian yields light-matter coupled singlet states on an equal footing , including both the singlet ground state and the excited singlet states.

\begin{align}
    &\langle \Phi_{i_\alpha}^{a_\beta} | \Delta\vec{\boldsymbol{\hat{d}}} | \Phi_{j_\alpha}^{b_\beta} \rangle =\nonumber\\
    &\begin{cases}
        \vec{d}_{a_\beta a_\beta} - \vec{d}_{i_\alpha i_\alpha},~~\mathrm{if}~~i = j~~\mathrm{and}~~a= b\\
       \vec{d}_{i_\alpha j_\alpha},~~\mathrm{if}~~i \ne j~~\mathrm{and}~~a= b \\
       \vec{d}_{a_\beta b_\beta},~~\mathrm{if}~~i = j~~\mathrm{and}~~a\ne b 
    \end{cases}.
\end{align}

\begin{figure}[t!]
    \centering
    \includegraphics[width=\linewidth]{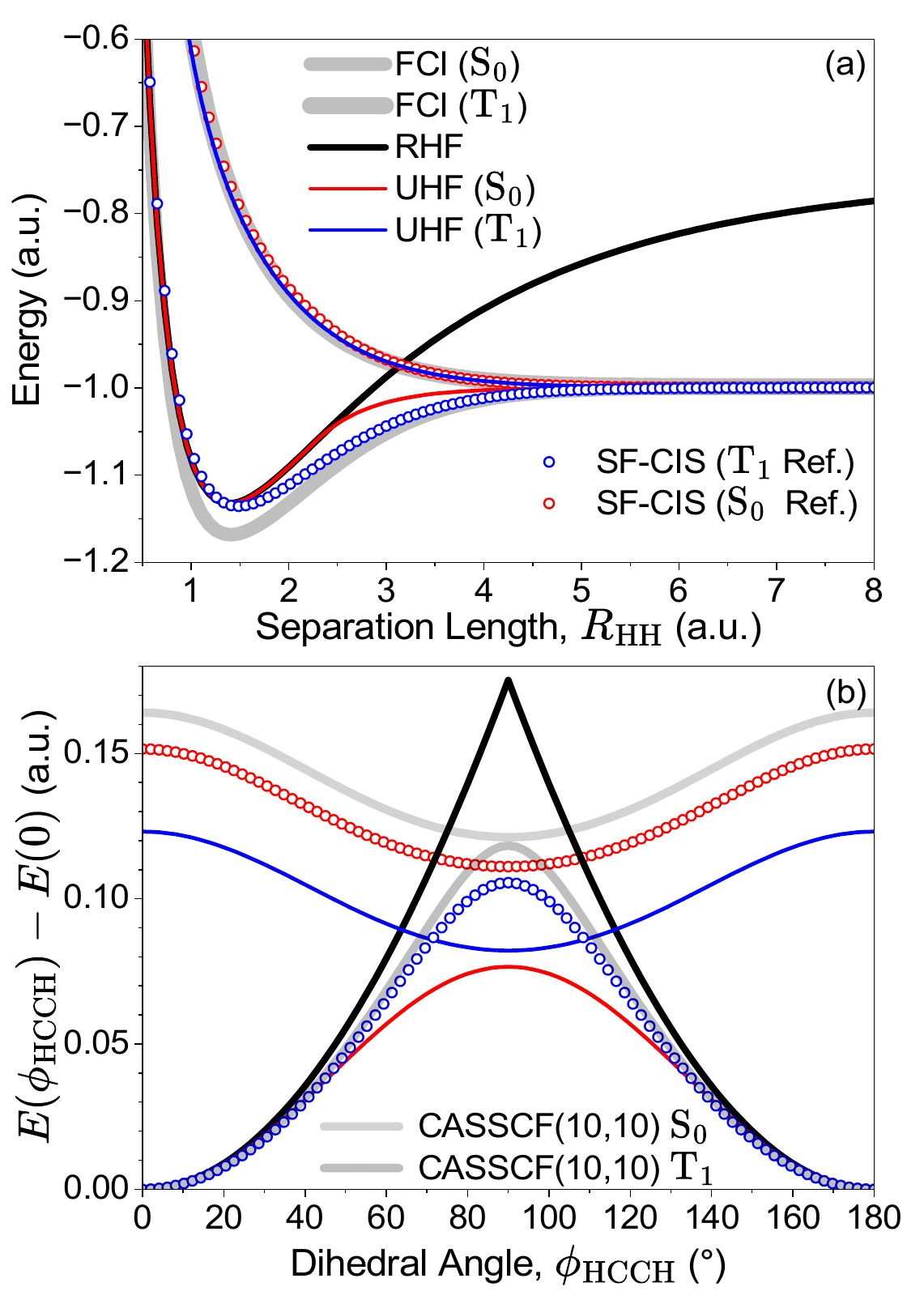}
    \caption{\footnotesize
    Singlet and triplet potential energy surfaces of the (a) H$_2$ dissociation coordinate $R_\mathrm{HH}$ and (b) ethylene dihedral angle $\phi_\mathrm{HCCH}$ at various levels of theory.
    }
    \label{FIG:H2_ETH_noCAVITY}
\end{figure}

\section{Results and Discussion}

To demonstrate the utility of proposed QED-SF-CIS method, we modeled several prototypical molecular systems. All calculations were performed using an in-house implementation interfaced with the PySCF package for evaluating atomic orbital integrals. Reference data were obtained from PySCF Full Configuration Interaction (FCI) and CASSCF(10,10), as shown in Figure~\ref{FIG:H2_ETH_noCAVITY}. All calculations in this work used the cc-pVDZ basis.

Figure~\ref{FIG:H2_ETH_noCAVITY} illustrates the performance of the spin-flip CIS approach for two simple chemical systems (hydrogen and ethylene molecules) outside the cavity. Figure~\ref{FIG:H2_ETH_noCAVITY}a shows the potential energy surfaces of the H$_2$ dissociation as functions of the internuclear coordinate, $R_\mathrm{HH}$. The spin-restricted Hartree-Fock (RHF, thick grey curve) shows the incorrect ionic limiting energy (and wavefunction) behavior provided by the standard Hartree-Fock theory. The spin-unrestricted HF (UHF) theory provides a singlet state (solid red curve) which has the correct limiting behavior of the energy. However, as is well known, the wavefunction is still incorrect and suffers from spin contamination beyond the Coulson-Fischer point.\cite{hait_beyond_2019} The UHF approach can also straightforwardly obtain the $M_s = \pm 1$ triplet states (solid blue curve). The spin-flip approach can take either the unrestricted singlet or triplet as the reference performing $\Delta M_s = +1$ (\textit{i.e.}, singlet-to-triplet) or $\Delta M_s = -1$ (\textit{i.e}, triplet-to-singlet) excitations, respectively. The spin-flip singlet state shown as blue circles (as obtained from the triplet reference shown as solid blue curve) deviates from the RHF result even at the minimum-energy geometry and tends toward the the full configuration interaction (FCI, solid black curve) result at increasing separation lengths. In contrast, the UHF singlet follows the RHF singlet until the Coulson-Fischer point, after which it tends toward the FCI result. Similarly, the spin-flip triplet state shown as red circles (obtained from the unrestricted singlet reference depicted as a sold red curve) closely matches the UHF triplet result, indicating only minimal improvement.

Figure~\ref{FIG:H2_ETH_noCAVITY}b presents the dihedral twisting of ethylene, $\phi_\mathrm{HCCH}$, a more complicated example where the ground state intersects with the excited state to form a conical intersection. In this case, we performed CASSCF(10,10) to obtain the ``exact'' reference result instead of FCI due to high computational cost. In this case, the RHF exhibits a cusp at $\phi_\mathrm{HCCH} = 90\degree$. At this geometry, the Aufbau-filling singlet ground state intersects with the two open-shell singlets and a doubly excited configuration. In fact, due to this degeneracy, the standard RHF variational SCF solution does not converge at this point. Thus, small displacements on this geometry are used to fill the curve near this point. The UHF singlet (solid red curve) allows for a correction to the energy which exhibits an avoided crossing, much like the CASSCF(10,10) reference (solid black curve). However, the barrier height (\textit{i.e.}, transition state energy) is significantly lower. The spin-flip singlet (blue circles) on the other hand provides a correction to the barrier height which becomes much closer to the CASSCF(10,10) result. However, it is still lower by $\sim 0.015$ a.u. ($\sim 400$ meV or $\sim 10$ kcal/mol). As before in the H$_2$ example, the triplet state is largely unaffected by the spin-flip methodology (compared to the UHF triplet) other than a systematic increase in energy concomitant with the shift in the singlet barrier height. 

\begin{figure}[t!]
    \centering
    \includegraphics[width=\linewidth]{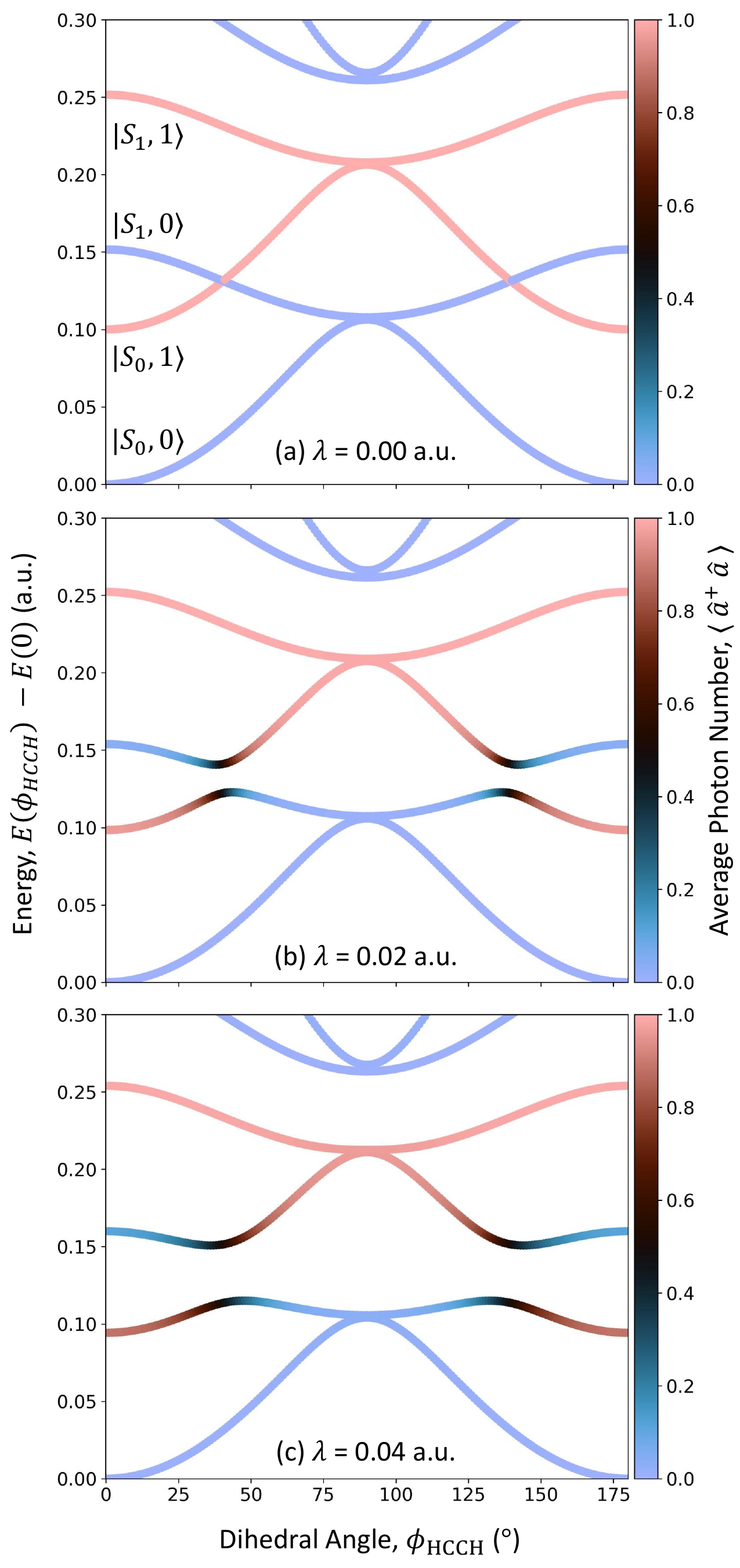}
    \caption{\footnotesize
    Potential energy surfaces of the four lowest-energy singlet polaritonic states S$_0$, S$_1$, S$_2$, and S$_3$ of the polaritonic system with $\omega_\mathrm{c} = 0.10$ a.u. (2.71 eV) for light-matter coupling strengths $\lambda$ = (a) 0.00, (b) 0.02, and (c) 0.04 a.u. The colorbar indicates the average photon character, $N_i = \langle \mathrm{S}_i |\hat{a}^\dagger \hat{a} | \mathrm{S}_i \rangle$ for each polaritonic state. At $\lambda = 0.0$ a.u., the non-interacting configurations are shown at $\phi_\mathrm{HCCH} = 0\degree$. Note that the character of the adiabatic states changes with nuclear configuration, $\phi_\mathrm{HCCH}$.
    }
    \label{FIG:ETH_EXCITED}
\end{figure}

Thus, Figure~\ref{FIG:H2_ETH_noCAVITY} highlights the usefulness  of the spin-flip approach in correcting the topology of the ground state potential energy surfaces, especially in cases where the ground state  strongly interacts with nearby excited states (\textit{i.e.}, the static correlation problem). These situations are known to fail in standard HF and CIS approaches -- as well as in DFT and TDDFT -- due to treating the reference and excited states on separate footing. This is resolved in the spin-flip approach since all states in the same spin-manifold are treated on an equal basis.

Continuing with the ethylene example for the remainder of this work (Figure~\ref{FIG:H2_ETH_noCAVITY}b), we now explore the cavity effects on the potential energy surfaces, computed at the spin-flip QED-CIS level. Figure~\ref{FIG:ETH_EXCITED} presents the strongly coupled excited states at various light-matter coupling strengths $\lambda$ = 0.00 (Figure~\ref{FIG:ETH_EXCITED}a), 0.02 (Figure~\ref{FIG:ETH_EXCITED}b), and 0.04 a.u. (Figure~\ref{FIG:ETH_EXCITED}c). The color of the curves indicates the average photonic character, $\langle \hat{a}^\dagger \hat{a} \rangle$. The cavity frequency is $\omega_\mathrm{c} = 0.10$ a.u. (2.72 eV). At zero light-matter coupling strength $\lambda$ = 0.0 a.u., at $\phi_\mathrm{HCCH} = 0\degree$, the non-interacting surfaces are energetically ordered as $|\mathrm{S}_0, 0\rangle$, $|\mathrm{S}_0, 1\rangle$, $|\mathrm{S}_1, 0\rangle$, and $|\mathrm{S}_1, 1\rangle$, followed by higher-lying states $|\mathrm{S}_2, 0\rangle$ and $|\mathrm{S}_3, 0\rangle$ (appearing in the figure near $\phi_\mathrm{HCCH} \in (40\degree,150\degree)$), which do not directly couple to the excited photon states in the energy range of interest.

At $\omega_\mathrm{c} = 0.10$ a.u., the resonance condition occurs, when the non-interacting $|\mathrm{S}_0,1\rangle$ and $|\mathrm{S}_1,0\rangle$ states are degenerate near $\phi_\mathrm{HCCH} \approx 40\degree$ (Figure~\ref{FIG:ETH_EXCITED}a). In this region, the two states interact to form polaritonic states which share the character of both the light and the matter, as shown by the colorbar of Figure~\ref{FIG:ETH_EXCITED}. At finite coupling strengths, $\lambda$ = 0.02  (Figure~\ref{FIG:ETH_EXCITED}b) and 0.04 (Figure~\ref{FIG:ETH_EXCITED}c)  a.u., a Rabi splitting $\Omega_\mathrm{R} \approx $ 0.02 (500 meV) and 0.04 (1000 meV) is observed at the resonance/crossing point. We emphasize that in realistic Fabry-Perot cavity designs, there are many molecules $N^6-N^{10}$ simultaneously coupled to the cavity mode. In this work, we consider only a single molecule with an increased light-matter coupling strength. It is well-known that the collective Rabi splitting scales with the number of molecules as $\Omega_\mathrm{R} \propto \sqrt{N}\lambda$. Therefore, to achieve the same Rabi splitting with $N$ molecules as that obtained with a single molecule, one would choose the corresponding single-molecule coupling strength to be $\lambda/\sqrt{N} \rightarrow \lambda$.

\begin{figure}[t!]
    \centering
    \includegraphics[width=\linewidth]{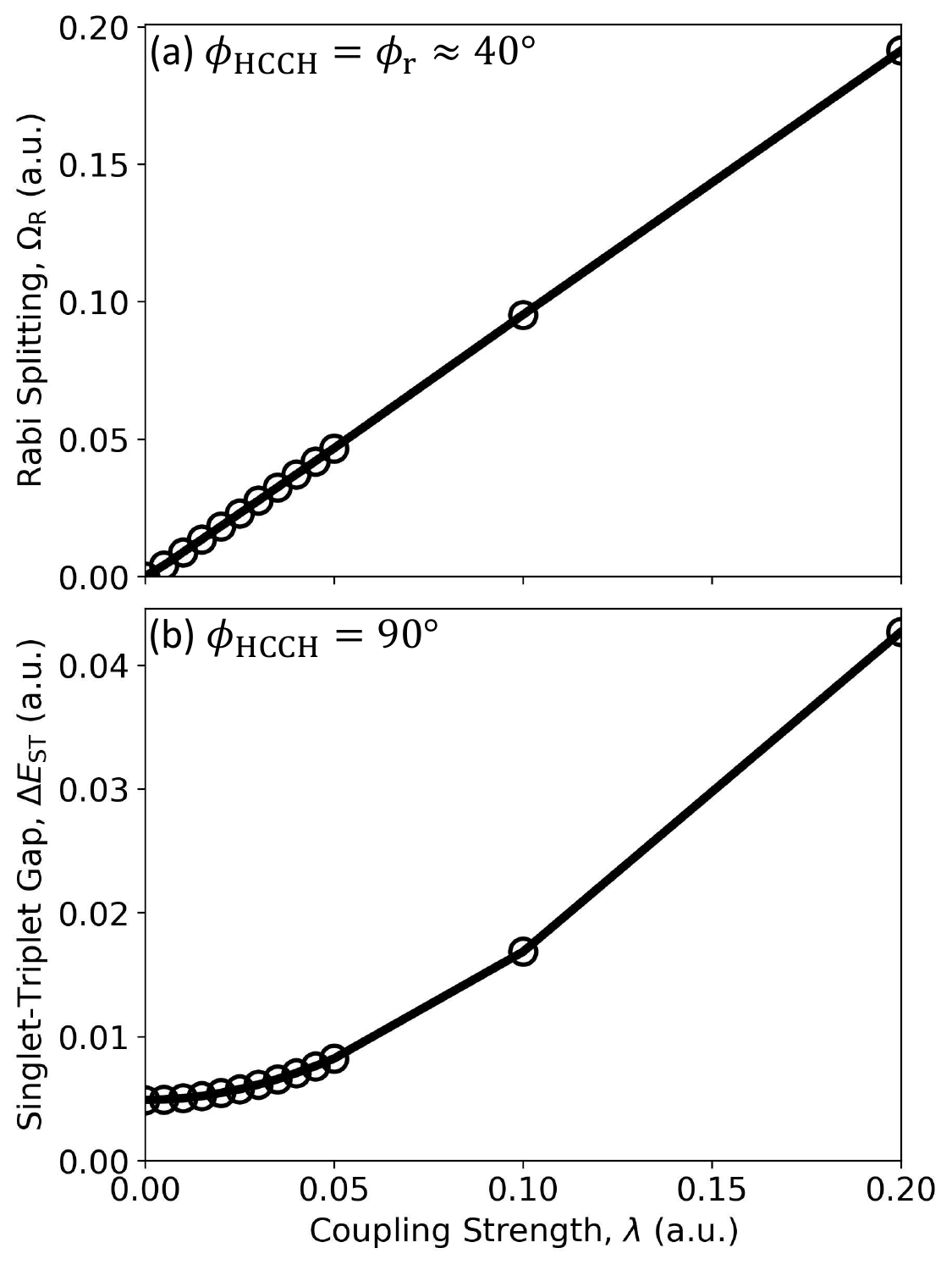}
    \caption{\footnotesize
    (a) Rabi splitting $\Omega_\mathrm{R} (\lambda,\phi_\mathrm{r}) = E_2(\lambda,\phi_\mathrm{r}) - E_1(\lambda, \phi_\mathrm{r})$ as a function of the light-matter coupling strength $\lambda$ for the dihedral which satisfies the resonance condition $\phi_\mathrm{r}$ ($E_2(\lambda = 0,\phi_\mathrm{r}) - E_1(\lambda = 0,\phi_\mathrm{r}) = 0$ at $\lambda = 0$ a.u.).. (b) Singlet-triplet ground state gap, $\Delta E_\mathrm{ST}(\lambda) = E^\mathrm{T}_0(\lambda,\phi_\mathrm{HCCH} = 90\degree) - E^\mathrm{S}_0(\lambda,\phi_\mathrm{HCCH} = 90\degree)$
    The cavity frequency was set to $\omega_\mathrm{c} = 0.10$ a.u. (2.71 eV).
    }
    \label{FIG:RABI_SPLITTING}
\end{figure}

Interestingly, with the introduction of the cavity degree of freedom, the excited state potential energy surface can be drastically modified by electric dipole-mediated interactions. This is important for nonadiabatic relaxation. As reported in the literature,\cite{hu_quasi-diabatic_2022,mandal_investigating_2019}  quantum dynamics simulations have shown that tuning the light-matter coupling strength $\lambda$ or cavity frequency $\omega_\mathrm{c}$ can alter the outcome of excited state chemical processes. This has also been shown theoretically for vibrational polariton ground state chemistries.\cite{schafer_shining_2022} At zero light-matter coupling (Figure~\ref{FIG:ETH_EXCITED}a), excitation to $|\mathrm{S}_1,0\rangle$ results in nearly 100\% rotation from $\phi_\mathrm{HCCH} = 0\degree$ to $\phi_\mathrm{HCCH} = 90\degree$ and beyond, since there are no barriers in the adiabatic potential energy surface. However, at finite coupling strengths, (Figure~\ref{FIG:ETH_EXCITED}b,c), a barrier appears in the excited state manifold (specifically in polaritonic adiabatic state $E_2(\phi_\mathrm{HCCH})$) due to the formation of polaritonic branches. This barrier prevents some or all dynamical trajectories, depending on the coupling strength) from reaching $\phi_\mathrm{HCCH} = 90\degree$.

Figure~\ref{FIG:RABI_SPLITTING}a shows the Rabi splitting at the resonance point $\phi_\mathrm{HCCH} \approx 40\degree$ as a function of the light-matter coupling strength. As expected, our results show this Rabi splitting to be a linear function of the coupling strength $\Omega_\mathrm{R} \propto \lambda$ in the range of coupling shown, $\lambda \in [0.0, 0.05]$ a.u.. This linear behavior does not persist at stronger coupling strengths, where interactions with additional electronic and Fock states through both the bilinear coupling and the DSE terms in Eq.~\ref{EQ:H_PF_CS}.\cite{yang_quantum-electrodynamical_2021}

We next explore the strongly coupled ground state. In this case, we move beyond the standard Fabry-Perot-type cavity toward few- or single-molecule plasmonic cavities, which exhibit cavity mode volumes comparable to the volume of the molecule itself. This enables single-molecule coupling strengths beyond what is possible in Fabry-Perot cavities and has been the subject of multiple theoretical works exploring cavity-modified ground states for chemical reactions.\cite{haugland_coupled_2020,haugland_intermolecular_2021,haugland_understanding_2025,vega_parameterized_2025,wang_investigating_2025,weight_cavity_2024} We emphasize that these ground state modifications are still in the exciton-polariton cavity frequency regime ($\omega_\mathrm{c} \sim 1$ eV), rather than in the vibrational polariton regime  ($\omega_\mathrm{c} \sim 0.1$ eV). It is also important to note that the ground state modifications are primarily dominated by the DSE term in Eq.~\ref{EQ:H_PF_CS}, rather than by the bilinear light–matter coupling term.\cite{rokaj_lightmatter_2018,weight_cavity_2024,weight_ab_2025-1,vega_parameterized_2025,wang_investigating_2025}

\begin{figure}[t!]
    \centering
    \includegraphics[width=\linewidth]{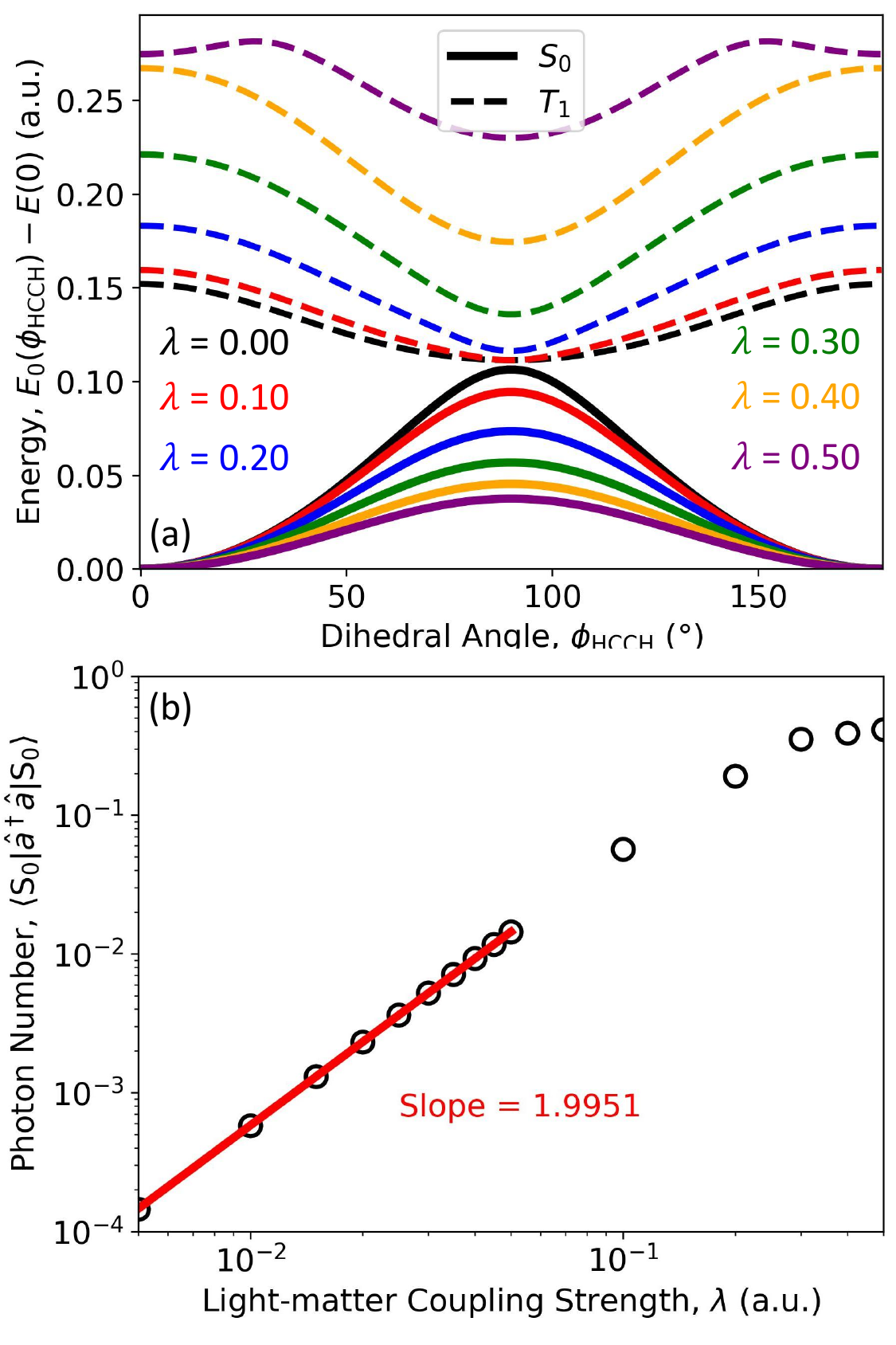}
    \caption{\footnotesize
    Potential energy surfaces of the singlet polaritonic ground state $S_0$ and the triplet ground state $T_1$ as functions of the dihedral angle $\phi_\mathrm{HCCH}$ of ethylene for a set of light-matter coupling strengths $\lambda$ = 0.00 (black), 0.10 (red), 0.20 (blue), 0.30 (green), 0.40 (orange), 0.50 (purple) with $\omega_\mathrm{c} = 0.10$ a.u. (2.71 eV). The linear fit (red) indicates the power of 1.9823 (nearly quadratic) for $\lambda < 0.05$ a.u.
    }
    \label{FIG:GROUND_STATE}
\end{figure}

Figure~\ref{FIG:GROUND_STATE}a shows the lowest-energy singlet S$_0$ (solid curves) and triplet T$_1$ (dashed curves) potential energy surfaces as functions of ethylene dihedral angle $\phi_\mathrm{HCCH}$ at various light-matter coupling strengths $\lambda$ = 0.0 (black), 0.10 (red), 0.20 (blue), 0.30 (green), 0.40 (orange), and 0.50 a.u. (purple). The singlet state is obtained by spin-flip ``down'' excitations of the UHF $\alpha\alpha$-triplet reference while the triplet excited state is obtained as spin-flip ``up'' excitations of the UHF singlet. The black solid and dashed curves ($\lambda = 0.0$ a.u.) are the same data as shown in Figure~\ref{FIG:H2_ETH_noCAVITY} as blue and red circles, respectively. 

As the light-matter coupling strength increases, the barrier height of the singlet ground state decreases, facilitating a higher probability of barrier crossing events. As pointed out in a previous work,\cite{weight_diffusion_2024} strong light-matter coupling acts to localize the electronic wavefunction since the bilinear term is simply the position operator $\hat{d} \propto \sum_i \hat{x}_i$, and the DSE term is the square of such an operator $\hat{d}^2 = \sum_{i}\hat{x}_i^2 + \sum_{ij}\hat{x}_i\hat{x}_j$ for each electron $i$ and $j$. We hypothesize that for the dihedral twist in ethylene, the barrier reduction arises from such effects and could be understood via the visualization of the wavefunction\cite{weight_diffusion_2024} or the difference density.\cite{weight_cavity_2024,haugland_coupled_2020,vega_parameterized_2025,wang_investigating_2025}

\begin{figure*}[t!]
    \centering
    \includegraphics[width=0.9\linewidth]{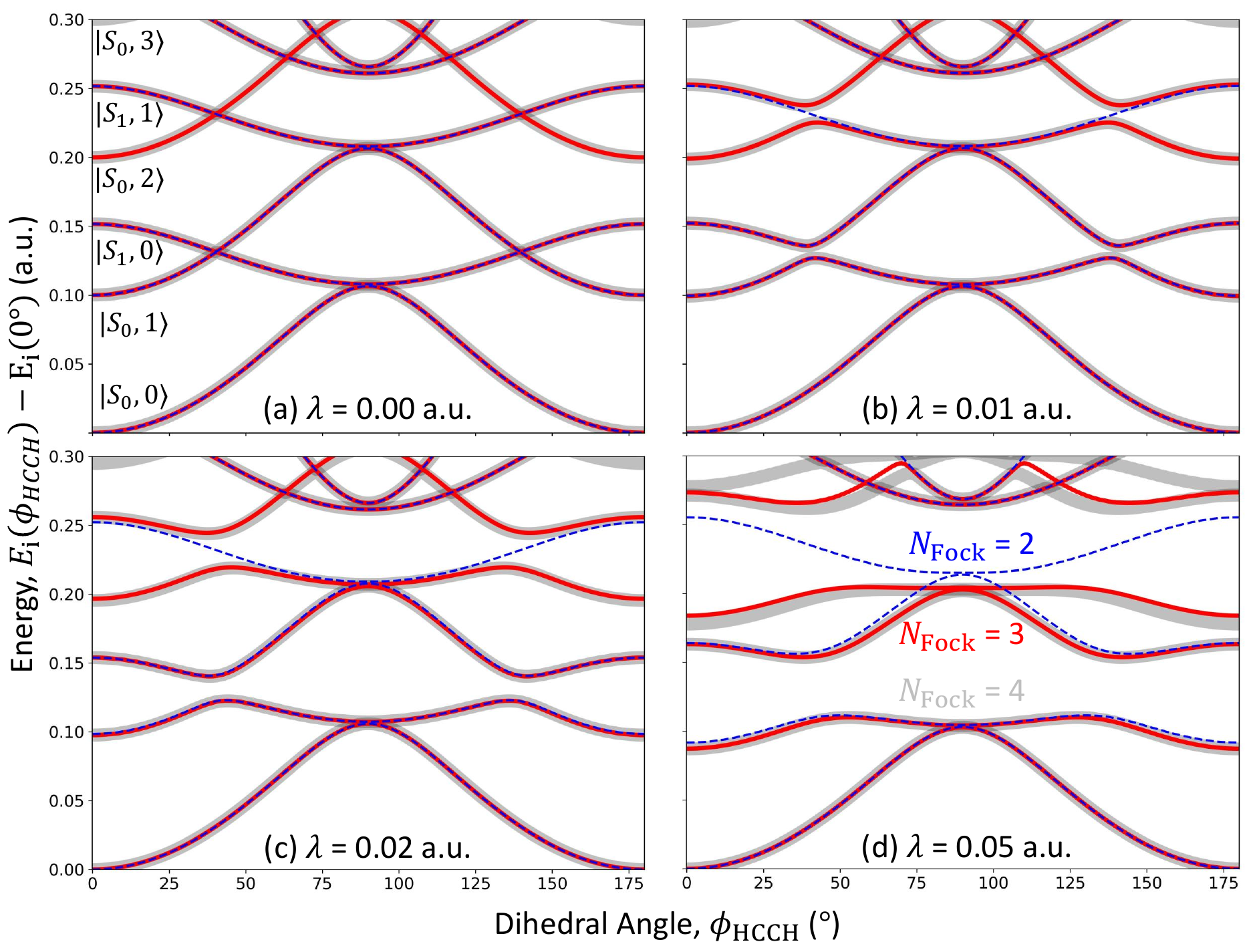}
    \caption{\footnotesize
    Potential energy surfaces along the ethylene dihedral twist angle $\phi_\mathrm{HCCH}$ at varying number of Fock states $N_\mathrm{Fock}$ = 2 (blue), 3 (red), and 4 (grey). Four values of light-matter coupling strength are shown: $\lambda$ = (a) 0.00, (b) 0.01, (c) 0.02, and (d) 0.05 a.u. At $\lambda = 0.0$ a.u., the non-interacting configurations are shown at $\phi_\mathrm{HCCH} = 0\degree$. Note that the character of the adiabatic states changes with nuclear configuration, $\phi_\mathrm{HCCH}$.
    }
    \label{FIG:NFOCK_CONVERGENCE_EXCITED}
\end{figure*}

Notably the singlet-triplet gap (see Figure~\ref{FIG:RABI_SPLITTING}b) also increases quadratically with increasing light-matter coupling strength. We believe that this is due to the excited photon state, \textit{i.e.}, $|\mathrm{S}_0,1\rangle$ interacting with the singlet ground state $|\mathrm{S}_0,0\rangle$ differently (with a different magnitude) than the excited photon state, \textit{i.e.}, $|T_1,1\rangle$ interacting with the triplet ground state $|T_1,0\rangle$. Recall that in the absence of spin-orbit coupling, as is the case here, the singlet and triplet manifolds are non-interacting, both inside and outside the cavity. The singlet-triplet gap (or any gap between adjacent spin states) is extremely important in metal-organic complexes, where determining the ground state multiplicity and, by extension, the corresponding equilibrium geometry, is an ongoing theoretical effort. These systems often exhibit near-degenerate spin-states stemming from rearrangement of the partially occupied d-orbitals which often relax to different geometries (\textit{e.g.}, tetrahedral, octahedral, etc.). The careful evaluation of their electronic structure, especially for heavy lanthanide compounds and materials, remains at the cutting edge of modern electronic structure theory. There are two possible directions in which cavity QED can directly contribute to this effort: (I) The electronic structure outside the cavity can be obtained in the limit that the light-matter coupling tends to zero. This is because, as seen here in Figure~\ref{FIG:RABI_SPLITTING}b, the energy gaps between spin states can be increased to reduce the static correlation in the system, allowing for single-reference approaches (including spin-flip approaches introduced here) to obtain superior accuracy. (II) The cavity can modulate the ordering of the spin states and allow for spin-conversion via strong coupling to the ground state. This opens the door for using materials that do not exhibit desirable properties in their intrinsic spin state (\textit{i.e.} outside the cavity). For example, coupling to quantized cavity fields may enable modification of the spin multiplicity of established organometallic catalysts, thereby altering their reactivity and functionality. The exploration of spin-flip approaches on metal-organic complexes within this light–matter framework will be the topic of future work. 

We further explore the expectation value of the number operator $\langle \mathrm{S}_0 | \hat{a}^\dagger \hat{a} | \mathrm{S}_0 \rangle$ for the coupled ground state as a function of the light-matter coupling strength $\lambda$, as shown in Figure~\ref{FIG:GROUND_STATE}b. The axes are log-log, whose linear function's slope indicates its power. Upon fitting to the lower coupling regime $\lambda \in [0.0, 0.05]$ a.u., we extract a power $1.9951$, indicating a quadratic function.\cite{weight_diffusion_2024,mctague_non-hermitian_2022} At larger couplings $\lambda > 0.05$ a.u., the function becomes sub-quadratic, in line with previous explorations at strong couplings.\cite{weight_diffusion_2024,mctague_non-hermitian_2022} The quadratic accumulation of photons in the ground state can be understood by assuming the following coherent state ansatz: $|\Psi\rangle \propto e^{-i\lambda(\hat{a}^\dagger + \hat{a})}|0\rangle$. Extracting the average photon number from this state results in $\langle 0 | e^{-i\lambda(\hat{a}^\dagger + \hat{a})} \hat{a}^\dagger \hat{a} e^{i\lambda(\hat{a}^\dagger + \hat{a})}| 0 \rangle = \lambda^2$. This implies that at larger couplings, the coherent state for the photon distribution among the Fock state is not retained.

\section{Converging Fock States}

At large coupling strengths $\lambda$, the convergence of the photon basis set becomes important, even when using the coherent state shift (Eq.~\ref{EQ:H_PF_CS}). We first generalize the QED-SF-CIS Hamiltonian (Eq.~\ref{EQ:H_QED_ab_SF_BLOCK}) to include two photon excitations $\{|0\rangle,|1\rangle\} \rightarrow\{|0\rangle,|1\rangle,|2\rangle\}$. This Hamiltonian can then be written as
\begin{widetext}
\begin{align}\label{EQ:H_QED_ab_SF_BLOCK_nFOCK_3}
    &\hat{H}^\mathrm{QED-SF-CIS}_{\alpha\rightarrow\beta}(N_\mathrm{Fock}=3) =
    \begin{bmatrix}
        \langle \Phi_\alpha^\beta | {\bf A}' | \Phi_\alpha^\beta \rangle& \sqrt{\frac{\omega_\mathrm{c}}{2}}\vec{\lambda}\cdot\langle \Phi_\alpha^\beta | \vec{\boldsymbol{\hat{d}}} | \Phi_\alpha^\beta \rangle & \textbf{0}\\
        \sqrt{\frac{\omega_\mathrm{c}}{2}}\vec{\lambda}\cdot\langle \Phi_\alpha^\beta | \vec{\boldsymbol{\hat{d}}} | \Phi_\alpha^\beta \rangle&\langle \Phi_\alpha^\beta | {\bf A}' | \Phi_\alpha^\beta \rangle + \omega_\mathrm{c} & \sqrt{2}\sqrt{\frac{\omega_\mathrm{c}}{2}}\vec{\lambda}\cdot\langle \Phi_\alpha^\beta | \vec{\boldsymbol{\hat{d}}} | \Phi_\alpha^\beta \rangle\\
        \textbf{0} & \sqrt{2}\sqrt{\frac{\omega_\mathrm{c}}{2}}\vec{\lambda}\cdot\langle \Phi_\alpha^\beta | \vec{\boldsymbol{\hat{d}}} | \Phi_\alpha^\beta \rangle & \langle \Phi_\alpha^\beta | {\bf A}' | \Phi_\alpha^\beta \rangle + 2\omega_\mathrm{c}
    \end{bmatrix}.
\end{align}
\end{widetext}
Note that the additional $\sqrt{2}$ appears in the couplings due to $\langle 2 | \hat{a}^\dagger | 1 \rangle = \langle 1 | \hat{a} | 2 \rangle = \sqrt{2}$. This Hamiltonian can be extended to arbitrary numbers of Fock states. Furthermore, the number of independent blocks does not increase, since the quantities $\langle \Phi_\alpha^\beta | {\bf A}' | \Phi_\alpha^\beta\rangle$ and $\sqrt{\frac{\omega_\mathrm{c}}{2}}\vec{\lambda}\cdot\langle \Phi_\alpha^\beta | \vec{\boldsymbol{\hat{d}}} | \Phi_\alpha^\beta \rangle$ need only be computed once. These matrix elements can then be diagonally modified with the cavity frequencies $n \omega_\mathrm{c}$ or scaled by $\sqrt{n}$ for the bilinear coupling terms. In general, the Hamiltonian can be written compactly with tensor products of the electronic and photonic subspace matrices and constructed via sparse matrix operations to preserve memory.

We first examine the convergence of the excited polaritonic states at weak-to-strong coupling in the Fabry-Perot-like cavities. Figure~\ref{FIG:NFOCK_CONVERGENCE_EXCITED} presents similar data as discussed previously in Figure~\ref{FIG:ETH_EXCITED}, now including three sets of curves corresponding to $N_\mathrm{Fock}$ = 2 (blue), 3 (red), and 4 (grey), for each coupling strength $\lambda$ = 0.00 (Figure~\ref{FIG:NFOCK_CONVERGENCE_EXCITED}a), 0.01 (Figure~\ref{FIG:NFOCK_CONVERGENCE_EXCITED}b),  0.02 (Figure~\ref{FIG:NFOCK_CONVERGENCE_EXCITED}c), and  0.05 (Figure~\ref{FIG:NFOCK_CONVERGENCE_EXCITED}d) a.u., respectively.
Figure~\ref{FIG:NFOCK_CONVERGENCE_EXCITED}a shows the non-interacting limit. Each basis configuration in labeled at $\phi_\mathrm{HCCH} = 0\degree$, noting that the ordering of the adiabatic states changes as a function of the nuclear configuration. 

Importantly, two new surfaces appear here that are absent in Figure~\ref{FIG:ETH_EXCITED}: $|\mathrm{S}_0,2\rangle$ and $|\mathrm{S}_0,3\rangle$. These states arise solely due to the expanded photonic basis, specifically the inclusion of the $|n =2\rangle$ and $|n = 3\rangle$ Fock states. As the coupling strength increases $\lambda$ = 0.01 a.u. and 0.02 a.u. (see Figure~\ref{FIG:NFOCK_CONVERGENCE_EXCITED}b,c), the Rabi splitting between $|\mathrm{S}_0,1\rangle$ and $|\mathrm{S}_1,0\rangle$ seen in Figure~\ref{FIG:ETH_EXCITED},  can be observed. Moreover, an additional Rabi splitting appears between $|\mathrm{S}_1,1\rangle$ and $|\mathrm{S}_0,2\rangle$, beyond the single-photon excited subspace considered previously. In the $N_\mathrm{Fock} = 2$ case (blue), the original $|\mathrm{S}_1,1\rangle$ is unimpeded across all coupling strengths since there is no coupling -- except from the DSE terms, which are small in these coupling strengths. 

\begin{figure}[t!]
    \centering
    \includegraphics[width=\linewidth]{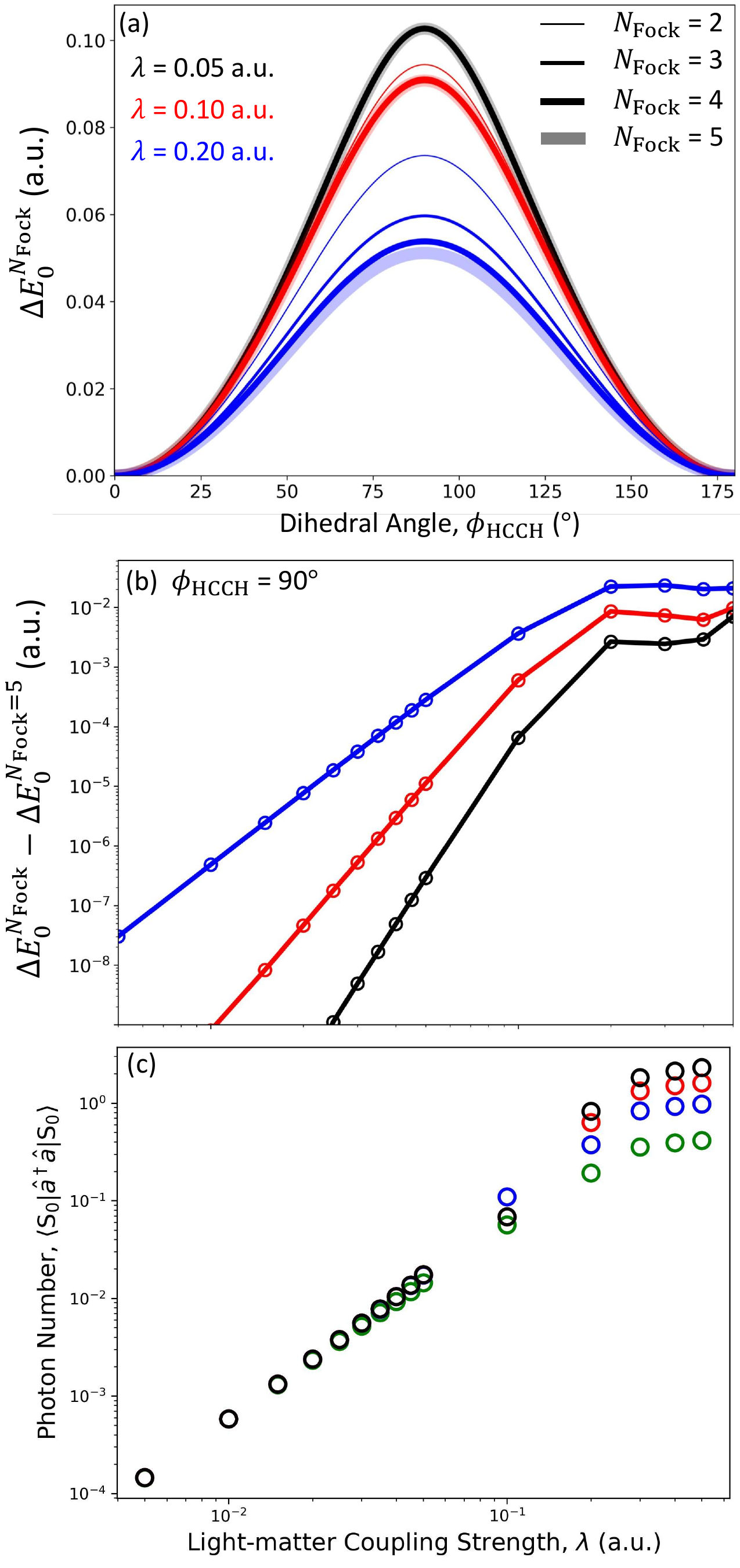}
    \caption{\footnotesize
    (a) Ground state potential energy surface, $\Delta E^{N_\mathrm{Fock}}_{0}(\phi_\mathrm{HCCH},0)$, for various number of Fock states $N_\mathrm{Fock}$ = 2, 3, 4, and 5 (lineiwdths) for three light0matter coupling strengths $\lambda$ = 0.05 (black), 0.10 (red), and 0.20 a.u. (blue). Here, $\Delta E^{N_\mathrm{Fock}}_{0}(\phi_\mathrm{HCCH},\phi_\mathrm{HCCH}') = E^{N_\mathrm{Fock}}_{0}(\phi_\mathrm{HCCH}) - E^{N_\mathrm{Fock}}_{0}(\phi_\mathrm{HCCH}')$. (b) Error in the ground state energy at $\phi_\mathrm{HCCH} = 90\degree$ (\textit{i.e.}, transition barrier energy) with respect to the number of Fock states, $\Delta E^{N_\mathrm{Fock}}_{0}(90\degree,0) - \Delta E^{N_\mathrm{Fock}=5}_{0}(90\degree,0)$. (c) Average photon number in the ground state at $\phi_\mathrm{HCCH} = 90\degree$ at various number of Fock states, $N_\mathrm{Fock}$ = 2 (green), 3 (blue), 4 (red), 5 (black). The $N_\mathrm{Fock} = 2$ results are identical to those shown in Figure~\ref{FIG:GROUND_STATE}b.
    }
    \label{FIG:NFOCK_CONVERGENCE_GROUND}
\end{figure}

At coupling strength $\lambda \le 0.02$ a.u., the $N_\mathrm{Fock} = 3$ and $N_\mathrm{Fock} = 4$ have nearly perfect agreement (except for the new $|\mathrm{S}_0,3\rangle$ state). At increased coupling $\lambda = 0.05$ a.u. (Figure~\ref{FIG:NFOCK_CONVERGENCE_EXCITED}d), the $N_\mathrm{Fock} = 3$ and $N_\mathrm{Fock} = 4$ start to diverge from one another. This is most evident for the higher-energy surfaces near $E \approx 0.25-0.30$ a.u., where the $|n = 4\rangle$ state becomes important. However, even for the first and second excited states, the agreement between $N_\mathrm{Fock} = 3$ and $N_\mathrm{Fock} = 4$ is reduced. As expected, the overall physics is recovered even with $N_\mathrm{Fock} = 2$ in the low-energy region of the spectrum.

Finally, we explore the convergence of the Fock state expansion for the strongly coupled ground state. Figure~\ref{FIG:NFOCK_CONVERGENCE_GROUND}a presents the potential energy surface for $N_\mathrm{Fock}$ = 2, 3, 4, and 5, where the thickness of the curves corresponds to the number of Fock states, and for three values of light-matter coupling strength $\lambda$ = (black) 0.05, (red) 0.10, and (blue) 0.20 a.u. At $\lambda = 0.05$ a.u., all curves are overlapping. At $\lambda$ = 0.10 and 0.20 a.u., the set of curves diverge, requiring more photonic excitations to converge the ground state. For $\lambda = 0.10$ a.u., the $N_\mathrm{Fock}$ = 3 and 4 are overlapping, implying that two photonic excitations are converged (visually, on this scale). For $\lambda = 0.20$ a.u., however, $N_\mathrm{Fock}$ = 3 does not agree with $N_\mathrm{Fock}$ = 4, differing by $\le 0.05$ a.u.

Figure~\ref{FIG:NFOCK_CONVERGENCE_GROUND}b presents the error in the transition state barrier height at $\phi_\mathrm{HCCH} = 90\degree$, compared to the reference barrier height of $N_\mathrm{Fock} = 5$, as a function of the light-matter coupling strength $\lambda$. The $N_\mathrm{Fock}$ = 2 (blue), 3 (red), and 4 (black) are shown on a log-log scale. At weak single-molecule couplings (Fabry-Perot-like cavities, $0 < \lambda < 0.05$ a.u.), the convergence is linear in the light-matter coupling strength $\lambda$ in the log-log scale with varying slopes, implying different powers, $\mathrm{error} \sim \lambda^n,~ 0 < \lambda <0.05$ a.u, for each set of basis configurations. At large coupling strengths (single- or few-molecule plasmonic cavities, $0.05 < \lambda < 0.5$ a.u.), the convergence is largely constant, except for at $\lambda = 0.5$ a.u., where the error begins to increase again. Realistically, $\lambda = 0.5$ a.u. likely exceeds what is what is currently achievable in state-of-the-art plasmonic cavities; therefore, there is little practical motivation to consider stronger coupling strengths. However, coupling strengths of $\lambda \le 0.3$ a.u. are within reach of frontier plasmonic cavity designs.\cite{weight_cavity_2024}

We then explore a convergence on the accumulated photon character in the ground state, as previously discussed in Figure~\ref{FIG:GROUND_STATE}b. This accumulation only arises due to the presence of the bilinear term, which mixed the character of adjacent photonic basis configurations, \textit{i.e.}, due to the $\hat{a}^\dagger + \hat{a}$ term. Including only the DSE term would not yield an accumulation of photonic character, but rather to mixing between electronic configurations mediated by the $\hat{d}^2$ term. It is also important to note that restricting the basis to a single photonic excitation limits any state's character (including the ground state) to have at most $\langle \hat{a}^\dagger \hat{a} \rangle = 1$.

Figure~\ref{FIG:NFOCK_CONVERGENCE_GROUND}c presents the average photonic character for $N_\mathrm{Fock}$ = 2 (green), 3 (red), 4 (blue), and 5 (black) included photonic configurations. At weak couplings ($\lambda < 0.05$ a.u.), there is minimal deviation of the curves, and $N_\mathrm{Fock} = 2$ provides a good approximation. At larger couplings ($0.05 < \lambda < 0.5$ a.u.), the average photonic character increases drastically with increased numbers of photonic excitations included in the basis. Note that the vertical axis is on a log-scale, so at $\lambda = 0.5$ a.u., the $N_\mathrm{Fock} = 5$ case exhibits $\langle \hat{a}^\dagger \hat{a} \rangle \approx 2-3$ while the $N_\mathrm{Fock} = 2$ case exhibits $\langle \hat{a}^\dagger \hat{a} \rangle \approx 0.5$. Thus, even with $N_\mathrm{Fock} = 5$, the photonic character is not expected to be fully converged. Again, $\lambda = 0.5$ a.u. represents an extreme coupling strength, slightly beyond what is currently accessible in state-of-the-art cavity designs. However, this regime may become experimentally accessible in the future and promises particularly intriguing physics. In such an ultrastrong-coupling limit, light and matter degrees of freedom become so strongly entangled that a  separation of particles' identities may no longer be meaningful, even in the ground state.

\section{Conclusions}

In this work, we have introduced an extension of the well-known spin-flip configuration interaction singles (SF-CIS) approach to treat strong statically correlated molecular materials in the presence of quantized cavity photons and their excitations, which we denote as QED-SF-CIS. We first introduce the bare SF-CIS formalism in the context of hydrogen dissociation and the ethylene dihedral twisting to showcase the power of the spin-flip approach in capturing the correct topologies of the ground state when interacting with nearly degenerate excited state configurations. We then explore the strongly coupled ethylene dihedral torsion inside an optical cavity, analyzing the effects in both a many-molecule Fabry-Perot-like cavity  (large mode volume, small single-molecule couplings) as well as a single- or few-molecule plasmonic cavity (low mode volume, large single-molecule couplings). Finally, we extend the intrinsic QED-SF-CIS approach, which includes only a single excitation in either degree of freedom (\textit{i.e.}, electron or photon), to include multiple photonic excitations, enabling improved convergence in the photonic space.

We note that spin-preserving linear response in the presence of cavity QED has been developed previously in a variety of implementations.\cite{flick_ab_2020,yang_quantum-electrodynamical_2021,weight_investigating_2023,mctague_non-hermitian_2022} Here, we extend this knowledge toward strong static correlation by introducing spin-non-preserving (\textit{i.e.}, spin-flip) excitations. Our formulation enables the description of both ground and excited states of the hybrid light–matter system on an equal footing. We further generalize this approach to include multiple excited photonic configurations, which are required in the strong coupling limit. We emphasize that the SF-CIS approach is not perfect and suffers from spin-contamination analogous to well known limitations of unrestricted HF theory. However, there are well-established approaches\cite{li_spin-adapted_2011,zhao_spin-adapted_2025,casanova_spin-flip_2008,sears_spin-complete_2003} provide systematic pathways to overcoming spin contamination.  The integration of these techniques within the QED-SF framework will be the focus of future work.

We anticipate that this approach will be broadly applicable in quantum dynamics simulations, particularly when combined with analytic gradients,\cite{yang_cavity_2022,li_machine_2024,zhang_non-adiabatic_2019,zhou_nuclear_2022,tichauer_multi-scale_2021,groenhof_tracking_2019} in systems exhibiting strong static correlation in both ground and excited states (\textit{e.g.} near S$_0$ and S$_1$ conical intersections). The method is also expected to be especially valuable for strongly coupled metal-organic compounds and complexes, where partially filled d-orbitals give rise to pronounced correlation effects. Since the SF-CIS approach provides access to both singlet and triplet manifolds within a unified framework, the spin-conversion in organometallic complexes can be explored inside the cavity. Here the strong coupling may bias the system toward a previously unfavorable molecular configuration and corresponding spin multiplicities that are inaccessible outside the cavity. 

These prospects are particularly compelling for heavy-metal systems, including lanthanides and actinides, where strong electronic correlation plagues electronic structure calculations. Strong coupling to the cavity may also alleviate some of these difficulties by lifting near-degeneracies among spin configurations through light–matter hybridization. It is our hope that this exploration lays the groundwork for future many-body approaches capable of accurately and efficiently describing strongly coupled light–matter systems in regimes where both electronic correlation and photonic effects are essential.

\vspace{1cm}

\section*{Acknowledgements}
 B.M.W. appreciates the support of the Director's Postdoctoral Fellowship at Los Alamos National Laboratory (LANL), which is funded by the Laboratory Directed Research and Development (LDRD) at LANL. LANL is operated by Triad National Security, LLC, for the National Nuclear Security Administration of the US Department of Energy (Contract No. 89233218CNA000001). The research was performed, in part, at the Center for Integrated Nanotechnologies (CINT), a U.S. Department of Energy, Office of Science user facility at LANL. Computing resources were provided by the LANL Institutional Computing (IC) Program. Approved for unlimited release (LA-UR-26-22085).





\bibliography{main} 
\bibliographystyle{rsc} 

\end{document}